\documentclass[onecolumn,prd,showpacs]{revtex4}
\usepackage{graphicx}
\usepackage{amssymb}
\begin{document}

\title{\bf A New Effective Potential for Deuteron}
\author{Taha Koohrokhi}
\email{t.koohrokhi@gu.ac.ir}
\affiliation{Department of Physics, Faculty of Sciences, Golestan University, Gorgan, Iran}

\author{Sehban Kartal}
\email{sehban@istanbul.edu.tr}
\affiliation{Istanbul University, Department of Physics, 34000, Istanbul, Turkey}

\begin{abstract}
We calculate for the first time the static properties of the deuteron, within the framework of supersymmetric quantum mechanics, analytically. A new effective potential and its partner are derived from a superpotential so that all parameters are fitted by the experimental data. An analytical expression is obtained for the deuteron wave function and contributions of the orthogonal $^{13}S_{1}$ and $^{13}D_{1}$ states are determined, explicitly. Compared to one pion exchange, the superpotential produces an electrostatic as well as two pion exchange terms for the potential. The saddle point radius of the potential and the maximum of the wave function are linearly proportional.
\end{abstract}

\maketitle
%
\section{Introduction}
Deuteron as the simplest nucleus consists of two nucleons. Study of deuteron provides useful information about static nucleon-nucleon (NN) interaction. The one pion exchange potential (OPEP) is an extended version of Yukawa potential that dominates in nucleon spacings above 3 fm, and is reasonable for spacings above 2 fm. However, the theoretical and experimental studies have shown that the nuclear force is not just a matter of an exchange of single pion \cite{Bertulani2007}. In addition, the exact study of NN interaction requires a more fundamental theory. Nevertheless, with good approximation, at long ranges, nucleons can still be considered structureless particles and OPEP is fitting for describing the NN interaction.

In general, NN interaction has been studied based on several main groups: Quantum Chromo
Dynamics (QCD) \cite{Myhrer1988629,Ping2009024001,Huang2018074018}, lattice QCD \cite{Beane2006012001,Ratti2018,Ishii2018434}, Effective Field Theory (EFT) \cite{RevModPhys811773}, Chiral Effective Field Theory (EFT) \cite{MACHLEIDT20111,PhysRevLett115122301,PhysRevC99024003}, Chiral Perturbation Theory (CHPT) \cite{ENTEM200293,PhysRevC91014,PhysRevC99024004}, Boson Exchange (BE) models \cite{SCHIERHOLZ1972335,PELAEZ20161,REUBER1996243}, Mean Field Theory (MFT) \cite{PTP1131009,Naghdi2014,Naghdi2014924} and phenomenological NN potentials \cite{PhysRevC63024001,PhysRevC492950,PhysRevC5138}. In most of the models, potentials have quite complicated structures and are described by many parameters. Indeed an efficient theory is a theory that, while having simple calculations and reproduction of expected values, has good insights, predictions and straightforward to be developed.

Supersymmetry (SUSY) was originally conceived within the quantum field theory as a means to unify the mathematical treatment of bosons and fermions \cite{COOPER1995267,WITTEN1981513}.
In this regard, supersymmetric quantum mechanics (SUSY QM) is a development of quantum mechanics that introduces new concepts such as superpotential, partner potentials, Hamiltonians hierarchy and shape invariant potentials.
The mathematical strategies in SUSY QM not only solve many problems algebraically but also classify potentials to different categories and determine criteria for them. Furthermore, approximation methods in SUSY QM provide more accurate results than those in conventional quantum mechanics \cite{Gangopadhyaya2017}. The theoretical successes of SUSY have caused its applications rapidly have been extended into other branches of physics and mathematics, i.e., supersymmetric quantum chromodynamics (SQCD) \cite{PhysRevD98085013} as well as nuclear physics \cite{Liang2016}, by unification of fermionic and bosonic fields to a superfield.

This paper is the first study of deuteron static properties by SUSY QM. Given the excellent results achieved, this approach can be developed straightforwardly considering more details of the interaction. Despite the simplicity of the calculations presented in this study, the proposed superpotential not only describes well the long ranges of the interaction but also gives notable results for intermediate and short ranges. The new attitude presented in this paper has introduced a new effective potential for deuteron that is also applicable for two-body interactions such as diatomic molecules.

 \section{Potential and Superpotential}
\subsection{OPEP}
The OPEP is the potential derived from meson theory in the
treatment of the NN system \cite{PTP16455}, is,
\begin{equation}
V_{OPEP}(r)=V_{C}(r)+S_{12}V_{T}(r)
\end{equation}
where $r$ is equal to the length of
the vector $\bf{r}$ connecting the two nucleons and $S_{12}$ is tensor operator. The first term in the OPEP is the central potential,
\begin{equation}
V_{C}(r)=V_{0}(\mathbf{\tau}_{1}.\mathbf{\tau}_{2})(\mathbf{\sigma}_{1}.\mathbf{\sigma}_{2})\frac{e^{-r/R}}{r}
\end{equation}
where $R=\frac{\hbar}{m_{\pi}c}$ is the typical range of the
nuclear force and $m_{\pi}$ is pion mass. The
neutron-proton interaction involves the exchanges of
both the neutral ($\pi^{0}$) and charged ($\pi^{\pm}$) pions.
For this reason, we employ the averaged-pion mass $\overline{m}_{\pi}=\frac{1}{3}(m_{\pi^{0}}+2m_{\pi^{\pm}})$ (Table 1) \cite{Babenko201758}. The dot products, $\mathbf{\tau}_{1}.\mathbf{\tau}_{2}$ and $\mathbf{\sigma}_{1}.\mathbf{\sigma}_{2}$, indicate the
isospin and spin dependencies of the potential, respectively.
The second term in the OPEP is called a tensor potential consisting a radial function and a tensor operator. Its radial part is,
\begin{equation}
V_{T}(r)=V_{0}(\mathbf{\tau}_{1}.\mathbf{\tau}_{2})\left[1+3\left(\frac{R}{r}\right)+3\left(\frac{R}{r}\right)^{2}\right]\frac{e^{-r/R}}{r}
\end{equation}
so that,
\begin{equation}
V_{0}=\frac{g^{2}\hbar c}{3}\left(\frac{\hbar c}{2MR}\right)^{2}
\end{equation}
where $M$ is nucleon mass and $g^{2}$ is an empirical constant. We use the mean value as, $\overline{g}^{2}_{\pi}=\frac{1}{3}(g^{2}_{\pi^{0}}+2g^{2}_{\pi^{\pm}})$ (Table 1) \cite{Babenko201758}. Experimental measurements for total spin-parity of deuteron give $J^{\pi} = 1^{+}$  [31]. The parity conservation and addition angular momenta rules
indicate that the ground state of the deuteron wave function contained only two $^{13}S_{1}$ ($T=0, S=1, L=0$ and $J=1$) and $^{13}D_{1}$  ($T=0, S=1, L=2$ and $J=1$) states.
With this consideration, the OPEP breaks the Schr\"{o}dinger equation into the two coupled equations \cite{Garcon2001},
\begin{equation}
\left\{ \begin{array}{ll}
\left\{\frac{\hbar^{2}}{2m}\frac{d^{2}}{dr^{2}}+E_{0}-V_{C}\right\}u(r)=\sqrt{8}V_{T}(r)\omega(r),\\
\left\{\frac{\hbar^{2}}{2m}\left(\frac{d^{2}}{dr^{2}}-\frac{6}{r^{2}}\right)+E_{0}+2V_{T}(r)-V_{C}\right\}\omega(r)=\sqrt{8}V_{T}(r)u(r),
\end{array}
\right.
\end{equation}
where $E_{0}$ is energy ground state of deuteron, $m$ is the reduced mass of proton-neutron system and $u(r)$ and $\omega(r)$
are the radial wave functions of $^{13}S_{1}$ and $^{13}D_{1}$ states, respectively. Concerning Eq. (5), $u(r)$ and $\omega(r)$ have different asymptotic behavior at large distances, because of the centrifugal potential. Furthermore, at short distances the same centrifugal barrier guarantees (at least for non-singular potentials $V_{C}$ and $V_{T}$) that $u(r)$ is proportional to $r$ and $\omega(r)$ is proportional to $r^{3}$. Nevertheless, the OPEP has presented here only for comparison. The main idea of the paper that starts from subsection 2.3 does not relate to solving this equation.


\subsection{A Unifying Potential}
Now we assume that $u(r)$ and $\omega(r)$ are proportional linearly \cite{Nicholson1962},
\begin{equation}
\omega(r)= \xi u(r).
\end{equation}
For a linear combination of $^{13}S_{1}$ and $^{13}D_{1}$ components, the ground state wave function of deuteron may be written as \cite{Wong1999},
\begin{equation}
\psi_{0}(\textbf{r})=a_{S}\psi_{S}(\textbf{r})+a_{D}\psi_{D}(\textbf{r}),
\end{equation}
with the normalization condition,
\begin{equation}
a^{2}_{S}+a^{2}_{D}=P_{S}+P_{D}=1.
\end{equation}
The wave function $\psi_{0}(\textbf{r})=R_{0}(r)Y_{\ell,m_{\ell}}(\theta,\varphi)\chi(S)T(t)$ is the product of radial $R_{0}(r)=\frac{U_{0}(r)}{r}$, angular $Y_{\ell,m_{\ell}}(\theta,\varphi)$, spin $\chi(S)$, and isospin $T(t)$ terms, respectively. The orthogonality of non-radial parts requires that \cite{Garcon2001},
\begin{equation}
\left\{ \begin{array}{ll}
|a_{S}|^{2}=P_{S}=\int_{0}^{\infty}u^{2}(r)dr\\
|a_{D}|^{2}=P_{D}=\int_{0}^{\infty}\omega^{2}(r)dr=\xi^{2}\int_{0}^{\infty}u^{2}(r)dr.
\end{array}
\right.
\end{equation}
where in the last term we use assumption (6). As a result, if expansion coefficients are real, i.e. $a^{*}_{S}=a_{S}$ and $a^{*}_{D}=a_{D}$, then we have,
\begin{equation}
\left\{ \begin{array}{ll}
u(r)=a_{S}U_{0}(r)\\
\omega(r)=a_{D}U_{0}(r)
\end{array}
\right.
\end{equation}
where $\xi=\frac{a_{D}}{a_{S}}$.

By replacing Eq. (10) in Eq. (5), we have,
\begin{equation}
\left\{ \begin{array}{ll}
a_{S}\frac{\hbar^{2}}{2m}U^{\prime\prime}_{0}(r)=a_{S}\left\{V_{C}(r)-E_{0}\right\}U_{0}(r)+a_{D}\sqrt{8}V_{T}(r)U_{0}(r)\\
a_{D}\frac{\hbar^{2}}{2m}U^{\prime\prime}_{0}(r)=a_{D}\left\{\frac{\hbar^{2}}{2m}\frac{6}{r^{2}}+V_{C}(r)-2V_{T}(r)-E_{0}\right\}
U_{0}(r)+a_{S}\sqrt{8}V_{T}(r)U_{0}(r),
\end{array}
\right.
\end{equation}
by adding these two equations, we now have an unified equation as,
\begin{equation}
\frac{\hbar^{2}}{2m}\frac{U^{\prime\prime}_{0}(r)}{U_{0}(r)}=V_{U}(r)
\end{equation}
where the unifying potential is,
\begin{equation}
V_{U}(r)=\frac{\hbar^{2}}{2m}\left(-\frac{2m}{\hbar^{2}}E_{0}+\frac{6b}{r^{2}}+\alpha\frac{e^{-r/R}}{r}+\beta\frac{e^{-r/R}}{r^2}+\gamma\frac{e^{-r/R}}{r^3}\right)
\end{equation}
and constant coefficients are,
\begin{equation}
\left\{ \begin{array}{ll}
b=\frac{a_{D}}{a_{S}+a_{D}}\\
\alpha=\frac{2m}{\hbar^{2}}\left[\mathbf{\sigma}_{1}.\mathbf{\sigma}_{2}+\left(\sqrt{8}-2b\right)\right]V_{0}\mathbf{\tau}_{1}.\mathbf{\tau}_{2}\\
\beta=\frac{6m}{\hbar^{2}}\left(\sqrt{8}-2b\right)V_{0}\mathbf{\tau}_{1}.\mathbf{\tau}_{2}R\\
\gamma=\frac{6m}{\hbar^{2}}\left(\sqrt{8}-2b\right)V_{0}\mathbf{\tau}_{1}.\mathbf{\tau}_{2}R^{2}.
\end{array}
\right.
\end{equation}
The corresponding values are listed in Table 2.

  \subsection{Superpotential and Partner Potentials}
In SUSY QM, by definition the superpotential as logarithmic derivative of ground state wave function \cite{Gangopadhyaya2017},
\begin{equation}
W(r)=-\frac{\hbar}{\sqrt{2m}}\frac{d}{dr}\ln{U_{0}(r)},
\end{equation}
the Schr\"{o}dinger equation as a quadratic differential equation is reduced to the first-order differential equation, as follows,
\begin{equation}
W^{2}(r)\mp\frac{\hbar}{\sqrt{2m}}W^{\prime}(r)=V_{\mp}(r),
\end{equation}
where minus and plus signs are related to $V_{-}(r)$ and $V_{+}(r)$, respectively. This equation is known as Riccati equation and $V_{-}(r)$ and $V_{+}(r)$ which are connected by the superpotential are known as supersymmetric partner potentials.
We now introduce an OPEP-like superpotential as follows,
\begin{equation}
W(r)=\frac{\hbar}{\sqrt{2m}}\left(A-\frac{L+1}{r}+C\frac{e^{-r/R}}{r}+D\frac{e^{-r/R}}{r^2}\right).
\end{equation}
Therefore, the ground state wave function obtained by replacing this superpotential in Eq. (15) equals to,
\begin{equation}
U_{0}(r)=N\exp\left\{-Ar+(L+1)\ln(r)+\left(C-\frac{D}{R}\right)\Gamma(0,r/R)+D\frac{e^{-r/R}}{r}\right\},
\end{equation}
where the normalization constant $N$ is acquired by $\int_{0}^{\infty}|U_{0}(r)|^{2}dr=1$,
and $A$, $L$, $C$ and $D$ are parameters to be determined by deuteron ground state properties in the next section.
 The superpotential $W(r)$ generates partner potentials via Eq. (16) as,
\begin{eqnarray}
V_{\mp}(r)&=&\frac{\hbar^{2}}{2m} \Big{\{}A^{2}-\frac{\alpha_{\mp}}{r}+\frac{\beta_{\mp}}{r^{2}}
+\gamma_{\mp}\frac{e^{-r/R}}{r}+\lambda_{\mp}\frac{e^{-r/R}}{r^2}+\chi_{\mp}\frac{e^{-r/R}}{r^3}\nonumber\\
&&+\left(C^{2}+\frac{2CD}{r}+\frac{D^{2}}{r^{2}}\right)\frac{e^{-2r/R}}{r^{2}}\Big{\}},
\end{eqnarray}
where its parameters are,
\begin{equation}
\left\{ \begin{array}{ll}
\alpha_{\mp}=2A(L+1)\\
\beta_{-}=L(L+1)~,~\beta_{+}=(L+1)(L+2)\\
\gamma_{-}=\left(2A+\frac{1}{R}\right)C~,~\gamma_{+}=\left(2A-\frac{1}{R}\right)C\\
\lambda_{-}=2AD+\frac{D}{R}-C(2L+1)~,~\lambda_{+}=2AD-\frac{D}{R}-C(2L+3)\\
\chi_{-}=-2DL~,~\chi_{+}=-2D\left(L+2\right)
\end{array}
\right.
\end{equation}
The obtained values are listed in Table 2.

 \section{Parameters Determination}
In order to determine six constants $A$, $a_{S}$, $a_{D}$, $L$, $C$ and $D$, we need six equations provided by using the static properties of deuteron ground state, as follows:
\subsection{$A$}
According to the unbroken SUSY, the ground state energy of $V_{-}(r)$ should be zero. Hence, the constant term in Eq. (19) is proportional to the ground state energy as,
\begin{equation}
A=\pm \sqrt{\frac{-2mE_{0}}{\hbar^2}}=\pm 0.2316,
\end{equation}
here we use experimental value for deuteron ground state energy $E_{0}$ (Table 1) \cite{RevModPhys841527}.
The plus sign is a valid selection for $A$ because the minus sign does not satisfy asymptotic condition ($\lim\limits_{r\rightarrow\infty} |U_{0}(r)|^{2}\ll 1$) for a bound state.

\subsection{$a_{S}$ and $a_{D}$}
The deuteron electric quadrupole moment is determined from the wave functions \cite{Garcon2001},
\begin{equation}
Q=e\int_{0}^{\infty}\left\{\frac{\sqrt{2}}{10}u(r)\omega(r)-\frac{1}{20}\omega^{2}(r)\right\}r^{2}dr,
\end{equation}
by using Eq. (10), we have,
\begin{equation}
Q=1.56a_{D}\left(\sqrt{2}a_{S}-\frac{a_{D}}{2}\right)e,
\end{equation}
We choose to use the empirical value for the
deuteron quadrupole moment (Table 1). By solving simultaneous two linear Eqs. (23) and (8), the two coefficients $a_{S}$ and $a_{D}$ are determined.

\subsection{Effective Angular Momentum $L$}
The third term in $V_{-} (r)$ is the centrifugal potential that always appears when we deal with the spherical coordinate system.
The expectation value of the square of angular momentum $\langle \hat{L}^{2}\rangle$ with $|\psi_{0}\rangle$ (Eq. 7) is equal to $6\hbar^{2}P_{D}$. As a result, by the following equality,
\begin{equation}
\hbar^{2}L(L+1)=6\hbar^{2}P_{D},
\end{equation}
we find an effective angular momentum as,
\begin{equation}
L=\frac{1}{2}\left(-1\pm\sqrt{1+24P_{D}}\right).
\end{equation}
the plus sign is acceptable for a real angular momentum.

\subsection{C and D}
\subsubsection{Wave Function Maximum}If deuteron has at least one bound state, its wave function should have a maximum inside the potential well.
 By assuming the maximum probability takes place at $r=R_{M}$, we have,
\begin{equation}
W(R_{M})=0,
\end{equation}
for special case $R_{M}=R$ an analytical expression obtain,
\begin{equation}
R_{M}=\frac{1}{2A}\left(L+1-0.368C+\sqrt{(0.368C-L-1)^{2}-1.47AD}\right).
\end{equation}
However, form Eqs. (26) and (17), the general expression $C$ in terms of $D$ is as follows,
\begin{equation}
C=R_{M}\exp\left(\frac{R_{M}}{R}\right)\left(-A+\frac{L+1}{R_{M}}\right)-\frac{D}{R_{M}}.
\end{equation}

\subsubsection{Structure Radius}
The deuteron structure $r_{\textrm{str}}$ and charge $r_{d}$ radius, were recently determined to
use several Lamb shift transitions in muonic deuterium which in by three times more precision
than previous measurements (Table 1) \cite{Science2016,Hernandez2018377}. On the other hand, deuteron structure
radius  as a characteristic deuteron size is defined from wave function, theoretically \cite{Garcon2001},
\begin{equation}
\langle r^{2}\rangle_{\textrm{str}}=\frac{1}{4}\int_{0}^{\infty}\left\{u^{2}(r)+\omega^{2}(r)\right\}r^{2}dr
\end{equation}
By putting Eq. (10) in Eq. (29) and using Eq. (8), the following equation is obtained,
\begin{equation}
\int_{0}^{\infty}\left[rU_{0}(r)\right]^{2}dr=15.6~(\textrm{fm}^{2})
\end{equation}
Fig. (1) illustrates $D$ for different values of $R_{M}$ obtained by numerical integrations resulting from the replacement of Eq. (28) into Eq. (30). We fit an exponential function on the result as,
\begin{equation}
D=-3.02\exp\left(\frac{R_{M}}{0.54}\right)+4.65.
\end{equation}
By changing the $C$ and $D$ with $R_{M}$, the normalization constant also is changed. We have performed a similar process for it, the results of which are shown in Fig. (2), and the fit of its exponential function is as follows:
\begin{equation}
N=-0.006\exp\left(\frac{R_{M}}{1.49}\right)+0.01.
\end{equation}


\section{Results and Discussions}
In unbroken SUSY, the two quantum systems described by supersymmetric partner potentials (superpartners)
have the same energy spectra except for the ground state of $V_{-}(r)$ \cite{COOPER1995267}.
The ground state energy of $V_{+}(r)$ is equal to the first excited state of
$V_{-}(r)$ and as a result $V_{+}(r)$ has less one energy level than $V_{-}(r)$. From the Fig. (3), it is seen that $V_{+}(r)$ has not the any attractive well and thus
bound states. This implies deuteron that is described by $V_{-}(r)$ is a weakly bound nucleus without any bound excited states. Furthermore, the superpartners are not shape invariant, thus we cannot obtain analytical solution for $V_{+}(r)$ \cite{koohrokhi2021}.

The NN interaction is usually classified into three main regions \cite{Wong1999}.
At short separation distances $(r\leq$ 1 fm) that is so-called hard core, it is
repulsive due to Pauli exclusion principle of identical fermions and incompressibility
of nuclear matter \cite{WANG2018207}. Similar to Van-der-Waals force in diatomic molecules,
saturation property is due to particle exchange as well as strongly repulsive forces at short distances. It means that the nuclear force becomes repulsive when
the nucleons try to get too close together. On other hand, at the intermediate-range $(1\leq r \leq$ 2 fm), the NN potential well is attractive
and causes to creation bound states. Finally, the OPEP and centrifugal potential are dominated at the long-ranges $r\geq 2$ (fm). In addition, in the asymptotic region ($r\rightarrow\infty$), the potential vanishes due to the finite range of the nuclear force between nucleons.
Fig. (3) shows that the potential $V_{-}(r)$ satisfies expected behavior in whole three mentioned regions.
To make a comparison, the potentials $V_{U}(r)$ and superpotential $W(r)$ are also depicted in Fig. (3).
It can be seen that all potentials have the same asymptotic behavior $\lim\limits_{r\rightarrow\infty} V(r)\rightarrow 0$ at the long ranges.
Among them, only $V_{-}(r)$ satisfies the intermediate and short ranges conditions.

The ground state wave function of deuteron $U_{0}(r)$ (400X magnification) is plotted in Fig. (3).
At the short ranges, the wave function is dropped rapidly due to the repulsive core. The peak of the wave function is located
at the intermediate range, near the edge of the well. It is the evidence of a weakly bound state within the potential well.
Ultimately, the wave function decreases at the long ranges gradually.

The wave function peak radius $R_{M}$ is surprisingly proportional with the saddle point radius $R_{S}$ of the $V_{-}(r)$ (Fig. (4)). We fit a line on the resulted data as,
\begin{equation}
R_{S}=R_{M}-0.09.
\end{equation}
This means regardless of value of $R_{M}$, the maximum probability of the presence of particles takes place near the well edge where the potential concavity sign is changed.

Let us now compare the new potential $V_{-}(r)$ (created by the superpotential) with $V_{U}(r)$ (created by the OPEP).
 It is clear that $V_{-}(r)$ contains two more terms than $V_{U}(r)$, as follows,
\begin{equation}
\left\{ \begin{array}{ll}
-\frac{2A(L+1)}{r}\\
\left(\frac{C^{2}}{r^{2}}+\frac{2CD}{r^{3}}+\frac{D^{2}}{r^{4}}\right)e^{-2r/R}
\end{array}
\right.
\end{equation}
The first one is due to Coulomb potential and the second term is related two-pion exchange potential.
It should be noted that these terms do not add to $V_{-}(r)$  by hand but are produced by the superpotential.
Since $2A(L+1)>$0, the Coulomb term implies an electric dipole moment (EDM) for deuteron \cite{PhysRevD101086009}. A permanent deuteron EDM
can arise, because a CP-violating neutron-proton interaction can induce a small $^{13}P_{1}$ admixture in the deuteron wave function, which
should be considered. As seen from Fig. (3), although this fact is more prominent at the shorter distances, this term has negligible contribution than the other components of the potential. Therefore, it is not significant impact in the practical applications at low energies.

In numerous studies have been shown that one boson exchange potentials (OBEP) are much easier to analyze than multi-meson ones.
Therefore, in most models, the multi-pion processes have been considered as the exchange of one
combined boson, rather than of multiple pions. To describe the attractive forces in the intermediate range,
OBEP models need a roughly 600 MeV $0^{+}$ scalar boson. In fact, many OBEP
models use both a 500 MeV and a 700 MeV scalar boson. The existence of
such scalar resonances has never been accepted. In this range, two-pion exchanges dominate. In such
exchanges, two pions appear during the course of the interaction. The typical range is correspondingly
smaller than for one-pion exchanges. Two-pion exchanges are much more difficult to crunch out than one-pion
ones. However, the superpotential $W(r)$, produces a two-pion exchange term in a simple way.


\section{Conclusion}
The most famous phenomenological models of the NN interaction (e.g. CD-Bonn, Reid93
and AV18) are based on the exchange of bosons and have many free parameters to be fitted with the experimental data \cite{PhysRevC63024001,PhysRevC492950,PhysRevC5138}. In the all models, the depth of potential well is inversely proportional to potential width. Moreover, each channel has its specific potential.
At the present unified picture, in contrast, $V_{-}$ is inseparable to $^{13}S_{1}$ and $^{13}D_{1}$ parts. As a result, the potential $V_{-}$ is narrower and deeper than those obtained by the other models. In addition, the $^{13}D_{1}$ state probability, $P_{D}$ (2-5 $\%$), is close to the lower limit. These probabilities should be recalculated due to the small $^{13}P_{1}$ admixture.

Actually, the model has been presented in this paper is a phenomenological model obtained by the superpotential approach. This new attitude has introduced a new superpotential and a corresponding effective potential ($V_{-}(r)$) for deuteron, as follows,
\begin{equation}
\left\{ \begin{array}{ll}
W(r)=\frac{\hbar}{\sqrt{2m}}\left(A-\frac{L+1}{r}+C\frac{e^{-r/R}}{r}+D\frac{e^{-r/R}}{r^2}\right)\\
V_{\rm eff}(r)=\frac{\hbar^{2}}{2m}\Big{\{}A^{2}-\frac{2A(L+1)}{r}+\frac{L(L+1)}{r^{2}}
+\left(\gamma_{-}+\frac{\lambda_{-}}{r}+\frac{\chi_{-}}{r^2}\right)\frac{e^{-r/R}}{r}\\
~~~~~~~~~~~~~+\left(\frac{Cr+D}{r}\right)^{2}\frac{e^{-2r/R}}{r^{2}}\Big{\}}
\end{array}
\right.
\end{equation}

These satisfy different static properties of force between nucleons, such as repulsive core at the short range, attractive at the intermediate range, finite range, spin and isospin dependencies, central and tensor parts, EDM and one- and two-pion exchanges. Also, some static properties of deuteron are satisfied by the superpotential including unitary and normalization, wave functions, probabilities, binding energy, charge radius, quadrupole moment, existence a weakly bound state as well as the lack of any excited states. Nevertheless, many questions about NN interaction are still  unanswered, for instance, magnetic moment, aspect ratio, scattering length, effective range, phase shifts, locality and non-locality properties, energy and momentum dependencies of NN interaction, etc.


\section{Acknowledgment}
The present study was carried out during T.K's sabbatical stay at Istanbul University. Authors acknowledge the financial support from Golestan and Istanbul Universities

\begin{table}[tbp]
\caption{Static properties of deuteron.}
\begin{center}
\begin{tabular}{ |c|c|c|c|c|c|c|c|c|c|c|c| }
\hline \scriptsize{$E_{0}$ (MeV)} & \scriptsize{$\overline{m}_{\pi}$ (MeV/c$^{2}$)} &
\scriptsize{$\overline{g}^{2}_{\pi}$} & \scriptsize{$r_{str}$ (fm)} & \scriptsize{$r_{d}$ (fm)}
& \scriptsize{$Q$ (efm$^{2}$)} & \scriptsize{R (fm)}\\
\hline \scriptsize{-2.22456627(46)} &
\scriptsize{138.039006} & \scriptsize{14.14} & \scriptsize{1.97507(78)} &
\scriptsize{2.12562(78)} & \scriptsize{0.2859(3)} & \scriptsize{1.4295}\\
\hline
\end{tabular}
\end{center}
\end{table}

\begin{table}[tbp]
\caption{The obtained coefficients for the potentials and superpotential for $R_{M}=R$.}
\begin{center}
\begin{tabular}{ |c|c|c|c|c|c|c|c|c|c|c|c| }
\hline \scriptsize{$a_{S}$} & \scriptsize{$a_{D}$} &
\scriptsize{$P_{S}~\%$} & \scriptsize{$P_{D}~\%$} & \scriptsize{A (fm$^{-1})$} & \scriptsize{L} & \scriptsize{C}
& \scriptsize{D (fm)} & \scriptsize{N}\\
\hline \scriptsize{$\pm0.99049$} &
\scriptsize{$\pm0.13755$} & \scriptsize{98.10} & \scriptsize{1.89} &
\scriptsize{0.2316} & \scriptsize{0.1029}
 & \scriptsize{28.09} & \scriptsize{-37.16} & \scriptsize{0.082}\\
\hline  \scriptsize{$\alpha$ (fm$^{-1})$} & \scriptsize{$\beta$} & \scriptsize{$\gamma$ (fm)}
& \scriptsize{$\gamma_{-}$ (fm$^{-1})$} & \scriptsize{$\lambda_{-}$}
& \scriptsize{$\chi_{-}$ (fm)} & \scriptsize{$\gamma_{+}$ (fm$^{-1})$} & \scriptsize{$\lambda_{+}$}
& \scriptsize{$\chi_{+}$ (fm)}\\
\hline \scriptsize{-1.30} & \scriptsize{-4.03}
& \scriptsize{-5.76}
& \scriptsize{32.66} & \scriptsize{-77.08} &
\scriptsize{7.65} & \scriptsize{-6.64}
 & \scriptsize{-81.27} & \scriptsize{156.27}\\
\hline
\end{tabular}
\end{center}
\end{table}

\begin{figure}
  \includegraphics[width=0.95\textwidth]{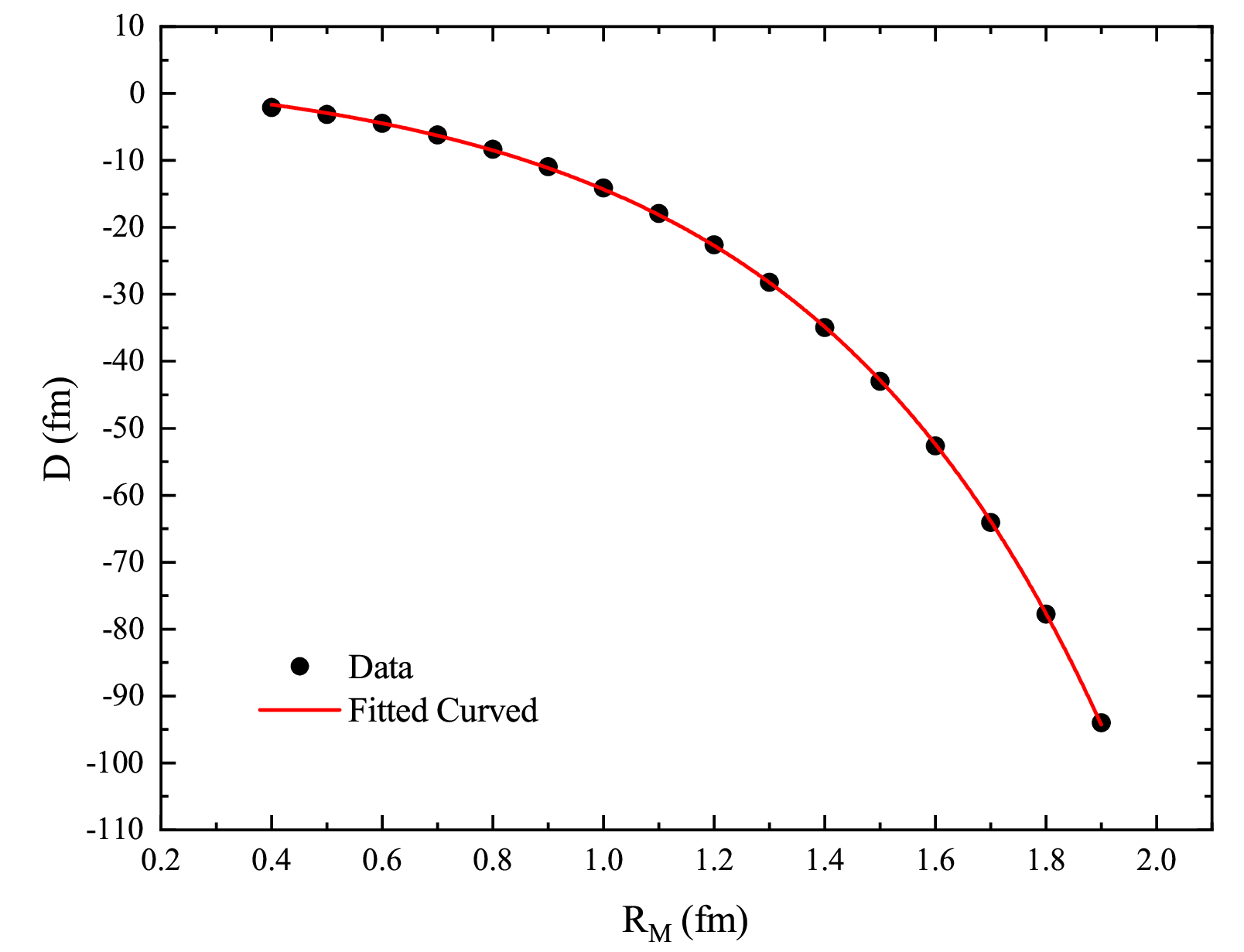}
\caption{$D$ versus $R_{M}$.} \label{fig:1}
\end{figure}

\begin{figure}
  \includegraphics[width=0.95\textwidth]{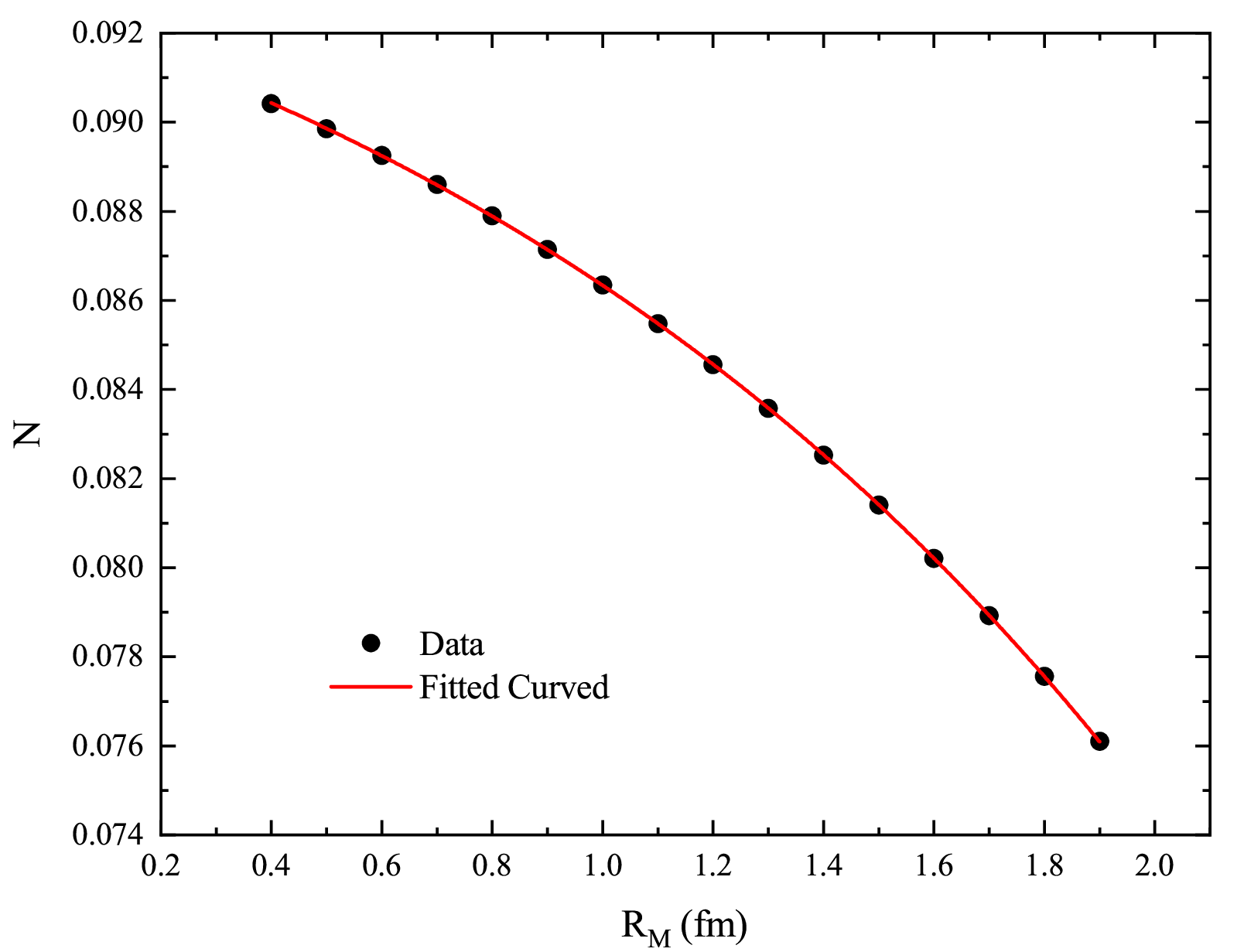}
\caption{Normalization constant versus $R_{M}$.} \label{fig:2}
\end{figure}

\begin{figure}
  \includegraphics[width=0.95\textwidth]{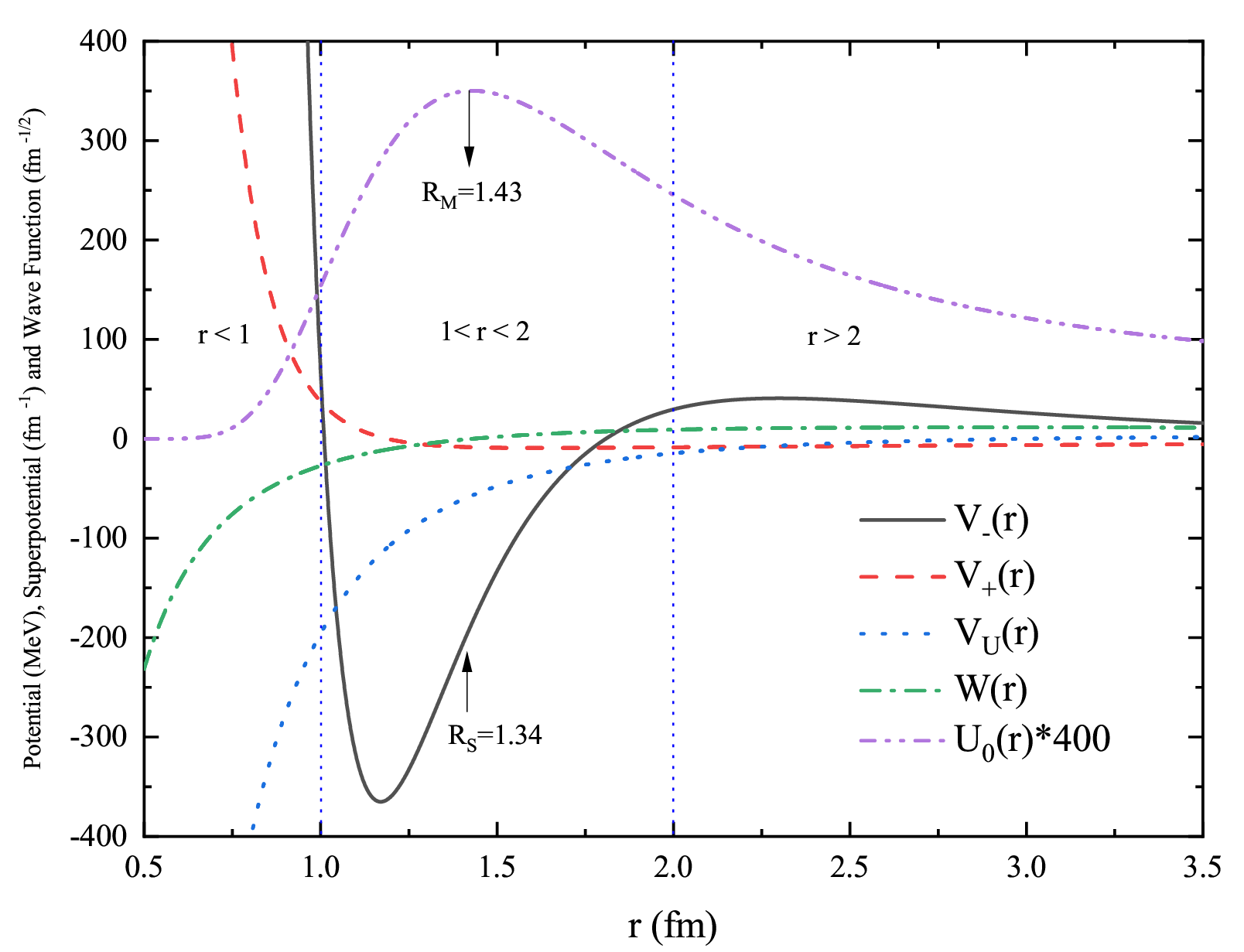}
\caption{This figure illustrates the partner potentials $V_{-}(r)$, $V_{+}(r)$, unifying potential $V_{U}(r)$, superpotential $W(r)$ and deuteron ground state wave function $U_{0}(r)$ for $R_{M}=R$.} \label{fig:3}
\end{figure}
\begin{figure}
  \includegraphics[width=0.95\textwidth]{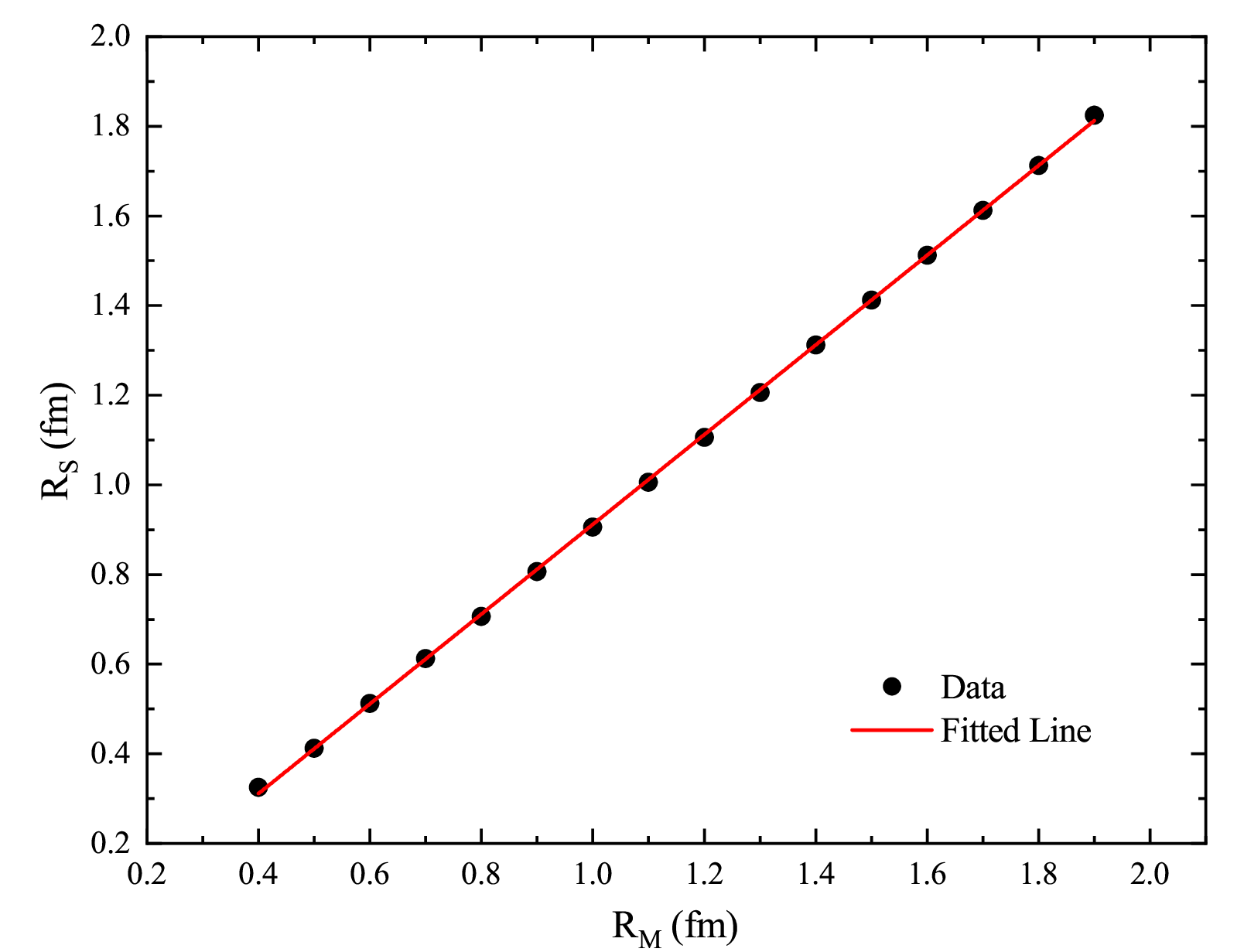}
\caption{Potential saddle point radius $R_{S}$ and maximum probability radius $R_{M}$.} \label{fig:4}
\end{figure}
\clearpage

\newpage

\bibliography{Deuteron}

\end{document}